\documentclass[12pt]{article}
\begin{document}
\begin{flushright}
UNILE-CBR3
\end{flushright}
\begin{center}
{\large \bf The Free-Energy of Hot Gauge Theories}
\end{center}

\vspace{0.5in}

\begin{center}
{Rajesh R. Parwani}

\vspace{0.5in}

{Departimento di Fisica, Universita' di Lecce,\\}
{and\\}
{Istituto Nazionale di Fisica Nucleare\\}
{Sezione di Lecce\\}
{Via Arnesano, 73100 Lecce, Italy.}

\vspace{0.25in}

\end{center}

\begin{abstract}
The total perturbative contribution to the 
free-energy of hot $SU(3)$ gauge theory is argued to lie 
significantly higher than the full result obtained by 
lattice simulations. This then suggests the existence of large 
non-perturbative corrections even at temperatures a few times
above the critical temperature.
Some speculations are then made on the 
nature and origin of
the non-perturbative corrections. The analysis is then carried out for
quantum chromodynamics, $SU(N_c)$ gauge theories, 
and quantum electrodynamics, leading to a conjecture and one more 
speculation.

\end{abstract}


\section{Introduction}
The most convincing evidence for a phase transition in thermal 
Yang-Mills theories
is provided by direct lattice simulation of the partition function,
\begin{equation}
Z= \mbox{Tr} e^{-\beta H} \, ,
\end{equation}
where $\beta =1/T $ is the inverse temperature. Over the years the 
lattice data for $SU(3)$ theory, the purely gluonic sector of 
quantum chromodynamics (QCD),
has become incresingly accurate, with
various systematic errors brought under control \cite{latt}. Figure (1)
shows the normalized free-energy density, $F=-T \ln Z /V$, 
of $SU(3)$ gauge theory  
taken from the first of Ref.\cite{latt}. Plots such as this have 
supported a picture of a low-temperature phase of glueballs 
melting above some critical 
temperature to produce a deconfined phase of
weakly interacting gluons: As the gluons are liberated the number of 
degrees of freedom increases causing the free-energy density to rise, while 
asymptotic freedom guarantees the gluons are weakly interacting at 
sufficiently high temperature.

Though the numerical data for  QCD is less accurate, due to 
technical difficulties in simulating fermions, the accumulated data
continues to support a phase transition. It is  generally believed 
that this is
a transition from a low-temperature hadronic 
phase to a high-temperature phase of quarks and gluons. This "quark-gluon
plasma" is the new phase of matter which experiments at Brookhaven and 
CERN hope to detect in the near future.

For the most part of this paper the focus will be on pure $SU(3)$ theory, 
since the accurate lattice data allows a direct comparison with theory.
Referring to Fig.(1), there is one feature which is ignored by some,
commented on by many and which has bothered a few. 
While there is little doubt
that at infinite temperature a description in terms of
gluons is tenable, this is less clear at
temperatures a few times the critical temperature, $T_c \sim 270 MeV$. 
For example, at $3T_c$, the curve lies $20\% $ below that of an ideal gas
of gluons. 

What is the origin of this large deviation ? Is it due  
to (i) perturbative
corrections to the ideal gas value, (ii) non-perturbative effects in the
plasma, (iii) an equally important combination of (i) and (ii) or, (iv)
is this an irrelevant question arising from an  
improper insistence of describing the high-temperature 
phase in terms of weakly coupled quasi-particles ?

Several viewpoints have been expressed in the literature. Some believe
that the deviation is mainly a non-perturbative correction to a 
gas of weakly coupled gluons and parametrize 
it in terms of a phenomenological "bag constant". 
Others have attempted a phenomenological 
description of the high-temperature phase 
in terms of generalized quasi-particles. For a discussion and 
detailed references to these phenomenological approaches 
see, for example, \cite{LH}.
On the other hand, a few have suggested that the best consistent 
description of the high-temperature 
phase might be in terms of novel structures \cite{detar, pis}. 
 
In order to help discern among the various possibilities, this paper will
focus on estimating the total perturbative contribution to the 
free-energy. It is important to first agree on some terminology so as to 
avoid confusion due to an overuse of some phrases in the literature.
The partition function 
depends on the Yang-Mills coupling $g$, and has a natural representation
as an Euclidean path-integral \cite{gpy}
\begin{equation}
Z(g) = \int D \phi \ e^{-\int_{0}^{\beta} d \tau \int d^3 x \ {\cal{L}}(\phi(x,\tau))} \,
\label{path}
\end{equation}
where $\phi$ collectively denotes the gauge and ghost fields, and $\cal{L}$ the
gauge-fixed lagrangian density of $SU(3)$ gauge theory. An expansion of this
path-integral, and hence the free-energy density, around $g=0$ 
leads to the usual Feynman perturbation theory
and contributions of the form $g^n$, with infrared effects 
occasionally generating logarithms multiplying the power terms, $g^n (ln(g))^m$.
These terms will be called perturbative.
What is invisible in a diagramatic expansion about $g=0$ are terms like
$e^{-1/g^2}$, associated for example with instantons \cite{gpy}. 
Such terms, 
which are exponentially suppressed as $g \to 0$, will be called non-perturbative.

Note that at non-zero temperature, odd powers of the coupling, such as $g^3$,
appear \cite{gpy}. These are perfectly natural and represent collective 
effects in the 
plasma. Though they sample interesting long-distance physics,
mathematically they fall into our definition of perturbative corrections.
Similarly, Linde \cite{lin} had shown that at order $g^6$
the free energy receives contributions 
from an infinite number of topologically distinct
Feynman diagrams.
Though the calculation of that contribution is difficult, it is possible
in principle \cite{bn}, 
and anyway does not qualify as 
a non-perturbative contribution according 
to the definition above. 

Following the heroic work of Arnold and Zhai, a completely analytical calculation
of the free-energy of thermal gauge theory to order $g^5$
has been obtained \cite{free,bn}. For $SU(3)$ gauge theory the result can be
summarized as follows,
\begin{eqnarray}
{F \over F_{0} } &=& 1 -{15 \over 4} \left({\alpha \over \pi}\right) + 30
\left({\alpha \over \pi}\right)^{3/2}
+ \left( 67.5 \ln\left({\alpha \over \pi}\right) + 237.2 -
20.63\ln\left({\bar{\mu} \over 2\pi T}\right) \right) \left({\alpha \over
\pi}\right)^2
\nonumber \\
&& \, -\left(799.2 - 247.5\ln\left({\bar{\mu} \over 2\pi T}\right) \right)
\left({\alpha \over \pi}\right)^{5/2} \, ,
\label{ym}
\end{eqnarray}
where $F_{0} = {-8\pi^2T^4 \over 45}$  is the contribution of non-interacting
gluons, $\alpha = g^2 / 4 \pi$, and $\bar{\mu}$ is the
renormalization scale in the $\overline{MS}$ scheme. 
Unfortunately (\ref{ym})
is an oscillatory, non-convergent, series even for $\alpha$ as small as $0.2$, 
which is close to the value of physical interest. A plot of (\ref{ym}) at different orders,
at the scale $\bar{\mu} =2 \pi T$ is shown in Fig.(2). The poor convergence of (\ref{ym})
does not allow a direct comparison of the perturbative results with lattice data. 
Furthermore, the result (\ref{ym}) is actually strongly dependent on the arbitrary
value of $\bar{\mu}$. Inspired by the relative success of Pade' resummation in other areas of
physics, Hatsuda \cite{H} and Kastening \cite{K} 
studied the Pade' improvement of the divergent series 
(\ref{ym}). Their conclusion was that the convergence could be improved, 
and the dependence on the scale $\bar{\mu}$ reduced. However they did not
attempt a direct comparison of their improved results with the lattice data,
though Hatsuda did conclude that for the case of four fermion flavours, 
the deviation
of the fifth-order Pade improved perturbative results from the 
ideal gas value 
was less than $10\% $. 

Not all seem to agree that a resummation of perturbative results 
as in \cite{H,K}
sheds sufficient light on the lattice data. For example,  
Andersen, {\it et.al.} \cite{and} and 
Blaizot, {\it et.al.} \cite{blaz} 
have abandoned the expansion
of the free-energy in terms of any formal parameter, but use instead 
gauge-invariance
as the main guiding principle to sum select classes of diagrams. Their 
low order results
seem to be close to the lattice data, but unfortunately 
because of the complexity
of the calculations and the absence of an expansion parameter,
it is not at all obvious what the magnitude of
'higher order' corrections is. A completely different
approach has been taken by Kajantie, {\it et.al}. In \cite{kaj}
an attempt has been made to numerically estimate the net contribution of 
long-distance effects, summarized in a dimensionally reduced effective
theory, to the free-energy density. As will be discussed later, the 
calculation of \cite{kaj} probably contains some of the 
non-perturbative effects defined
above but might miss out on some others.

This author believes that the declared
demise of information content in perturbative results such as 
(\ref{ym}) is premature. In Ref.\cite{rp1} a resummation
scheme was introduced to obtain an estimate of the total perturbative
contribution to the free-energy density of $SU(3)$ theory. The 
methodology of Ref.\cite{rp1} has been further developed and applied
to other problems in \cite{rp2,rp4}. In Secs.(2-5), an explicit
and improved discussion of the results in \cite{rp1} is given, 
leading to the conclusion that the {\it total} perturbative contribution 
to the free-energy density lies significantly above the 
full lattice data. In Sec.(6) I discuss the consequent magnitude of 
non-perturbative 
contributions, and speculate on their possible origin. Sec.(7) contains 
an analysis of the perturbative free-energy of generalized QCD, with
$N_f$ fermions, and a brief comparison with the available 
lattice data which is less precise. $SU(N_c)$ gauge theory is discussed in
Sec.(8), and an apparently universal relation noted. 
Sec.(9) considers quantum 
electrodynamics (QED)
and some speculations about its high-temperature phase. 
A summary and the conclusion is in Sec.(9).

\section{The Resummation Scheme}

The truncated perturbative expansion of the normalized free-energy density
can be written as 
\begin{equation}
\hat{S}_{N}(\lambda) = 1 + \sum_{n=1} ^{N}  f_{n} \lambda^{n} \, ,
\label{ser}
\end{equation}
where $\lambda = (\alpha / \pi)^{1 \over 2}$ is the coupling constant,
and where following \cite{H,K}, possible logarithms of the coupling
constant are absorbed into the coefficients $f_n$. The poor convergence of
(\ref{ser}) is obviously due to the large coefficients at high order. Indeed
such divergence of perturbative expansions is generic 
in quantum field theory and one expects the coefficients $f_n$
to grow as $n!$ for large $n $ \cite{order}. This leads to the   
introduction of the Borel transform
\begin{equation}
B_{N}(z) = 1 + \sum_{n=1}^{N} {f_{n} z^{n} \over  n!}
\label{borel}
\end{equation}
which has better convergence properties than (\ref{ser}). The 
series (\ref{ser}) may then be recovered using the Borel integral
\begin{equation}
\hat{S}_{N}(\lambda) = {1 \over \lambda} \int_{0}^{\infty} \ dz \ e^{-z/\lambda}
\ B_{N}(z) \, .
\label{lap}
\end{equation}
The logic of Borel resummation is to define the total 
sum $S(\lambda)$ of the perturbation expansion as the $N \to \infty$ limit 
of (\ref{lap}). This of course requires knowledge 
of $B(z) \equiv B_{\infty}(z)$ and the existence of the Borel integral.
Lacking knowledge of the exact $B(z)$, one 
therefore attempts to reconstruct
an approximation to $S$ by replacing the partial series $B_N(z)$ in (\ref{lap})
by a possible analytical continuation thereof. A simple and 
popular method to achieve this is to use Pade' approximants, leading therefore
to a Borel-Pade' resummation of the series (\ref{ser}). This method will 
be briefly discussed in Sec.(5). Here instead I will proceed as 
suggested by Loeffel \cite{L} and change variables in 
(\ref{lap}) through a conformal map.
For a positive parameter $p$, define
\begin{equation}
w(z) = { \sqrt{1 + pz} -1 \over \sqrt{1+pz} +1 }
\label{con}
\end{equation}
which maps the complex z-plane (Borel plane) to a unit circle in the $w$-plane. 
The inverse of (\ref{con}) is given by

\begin{equation}
z= { 4w \over p} {1 \over (1-w)^2 } \, .
\label{inv}
\end{equation}
The idea \cite{LZ} is to rewrite (\ref{lap}) in terms of the variable $w$. Therefore,
using (\ref{inv}), $z^n$ is expanded to order $N$ in $w$ and substituted into
(\ref{borel},\ref{lap}). The result
is
\begin{equation}
S_{N}(\lambda) = 1 + {1 \over \lambda} \sum_{n=1}^{N} {f_{n} \over n!} \left({ 4
\over p}\right)^n \, \sum_{k=0}^{N-n}
{ ( 2n+k-1)! \over  k! \ (2n-1)!}
 \int_{0}^{\infty} e^{-z/\lambda} \ w(z)^{(k+n)} \, dz \; ,
\label{rser}
\end{equation}
where $w(z)$ is given by (\ref{con}). Equation (\ref{rser}) represents
a highly nontrivial resummation of the original series
(\ref{ser}) \cite{rp2}. 
In the pioneering application of the Borel-conformal-map
technique in condensed matter physics by Le-Guillou and Zinn-Justin \cite{LZ}, 
the paramter $p$ was a fixed constant which determined the precise 
location of the instanton singularity at $z=-1/p$. In some more
recent QCD applications \cite{qcd},
the fixed constant $p$ determines the ultraviolet renormalon singularity
closest to the origin in the Borel plane \cite{ben}.

The novelty introduced in Ref.\cite{rp1} and futher developed in \cite{rp2}
was to consider $p$ as a variational parameter
determined according to the
condition
\begin{equation}
{\partial S_{N}(\lambda_o, p) \over \partial p} =
0 \, ,
\label{ext}
\end{equation}
where $\lambda_0$ is a convenient reference value. For the
problem at hand, because $f_2$ is negative, the solutions $p(N)$ to
(\ref{ext}) will be postions of minima \cite{rp2}. Denote the
value of (\ref{rser}) at $p=p(N)$ by $S_N(\lambda)$. Notice that although 
$p(N)$ is determined at the reference value $\lambda_0$, $S_N(\lambda)$ 
is defined for all $\lambda$. The reason why this is 
sufficient has been explained 
in \cite{rp2} and will be discussed further in the next section.

It must be stressed that, in general (see \cite{rp2}), 
the variational parameter $p(N)$ is not related to 
possible singularities of $B(z)$. Rather, it is determined according to
the extremum condition (\ref{ext}). Thus the presentation here 
is a slight departure from that in \cite{rp1} and represents 
the developments in Ref.\cite{rp2}. 

Sufficient information is now available to construct the resummed
approximants $S_N$ from $N=3$ up to $N=5$ in the next section. 
The approximant $S_1$ of course does not exist since $f_1=0$, while
$S_2$ cannot be constructed because no solution exists to equations such as
(\ref{ext}) at the first nontrivial order. For $N>2$, the
solutions $p(N)$ 
will be positions of global (local) minima if the sign of $f_N$
is positive (negative) \cite{rp2}.

\section{SU(3) : Resummation up to Fifth Order}

In order to make contact with lattice data which show a 
temperature dependent curve, one must use in (\ref{ser}) a 
temperature dependent coupling. Let us begin by using 
the one-loop running coupling defined by \cite{and, blaz},
\begin{equation}
\lambda(c,x) = {2 \over \sqrt{11 L(c,x)}} \, ,
\label{oneloop}
\end{equation}
where $L(c,x) = \ln((2.28 \pi c x )^2)$, $c=\bar{\mu} / 2 \pi T$ and $x=T/T_c$,
with $T_c \sim 270 MeV$ the critical temperature which separates the low and
high temperature phases \cite{latt,gupta}. 

Fixing first the reference values $c_{0}=1, \ x_{0}=1$,
which fixes the reference value of $\lambda_{0}$,
the results of (\ref{ext}) are: $p(3)= 3.2, \ p(4)=7.9, \ p(5) = 13.7$.
The curves for $S_N(\lambda)$ are shown in Fig.(3a) at the 
renormalization scale $c=1$. Notice the behaviour $S_5 > S_4 > S_3$
and how these all lie significantly above the lattice curve in Fig.(1).
The results do depend on the renormalization scale, denoted here by
the dimensionless parameter $c$. It has been suggested \cite{and,blaz}
that a suitable choice for such a parameter is $0.5<c<2$, corresponding to
$\pi T < \bar{\mu} < 4 \pi T$. Certainly this is the natural energy 
range for the high-temperature phase. Figure (3b) shows the
mild dependence of $S_5$ on the renormalization scale.

The results above were obtained by solving (\ref{ext}) at the point
$c_0=x_0=1$. Now consider changing the reference values to $c_0=1, \ x_0=3$,
that is, a more central value for the temperature. The solutions are:
$p(3)=3.2, \ p(4)=7.8, \ p(5)=13.4$. These values are hardly 
different from those above.
This is firstly due to the fact that (\ref{rser}) is a much slower
varying function of the coupling than the original divergent series.
Furthermore, for the  present problem, the coupling itself varies
slower than logarithmically with $c$ and $x$ (the $c$ and $x$ 
dependence of the coefficients $f_4$ and $f_5$ is also only logarithmic).
The curves for the re-optimized $S_N$ are essentially identical to those
shown in Fig.(3a,b), the difference being only at the fifth decimal point.
For example, the value of $S_5$ in Fig.(3a) at $x=3$ is $0.938684$, while that
for $S_5$ optimized at $x_0=3$ (and hence evaluated at p(5)=13.4), is 
$0.938672$ at $x=3$. This confirms the assertion in \cite{rp1} that the results
are quite insensitive to the exact reference values chosen to
solve (\ref{ext}).

We now proceed to test the sensitivity of the results to the approximation
used for the running coupling (\ref{oneloop}). The approximate
two-loop running coupling is given by \cite{and,blaz}
\begin{equation}
\lambda(c,x) = {2 \over \sqrt{11 L(c,x)}} \left( 1-{51 \over 121}
{\ln(L(c,x)) \over L(c,x)} \right)
\label{twoloop}
\end{equation}
with the symbols having the same meaning as before. In Fig.(4), the
one-loop running coupling (\ref{oneloop}) and the approximate two-loop
running coupling (\ref{twoloop}) are plotted at $c=1$ to show their
difference. At $x=3$, the value for the approximate 
two-loop coupling is about 
$20\%$ lower than the one-loop result. Nevertheless, because of the
above-mentioned property of the resummed series, we shall see that 
the final results
to do not shift dramatically. Using (\ref{twoloop}),
the solution of (\ref{ext}) at the reference point $c_0=1, x_0=3$
are: $p(3)=3.2, \ p(4)=7.6, \ p(5)=13.1$. The corresponding curves 
shown in Fig.(5a) have moved up slightly compared to those in Fig.(3a).
The "two-loop" value of $S_5(c=1,x=3) = 0.9473$ should 
be compared to the "one-loop" value $0.9387$ obtained above. 
The mild renormalization scale dependence of the new $S_5$ is shown
in Fig.(5b).

In summary, it has been demonstrated that the resummed approximants
$S_3, \ S_4, \ S_5$ all lie significantly above the lattice data and
satisfy the monotonicity condition $S_5 > S_4 > S_3$. The result is 
insensitive to the reference value used to solve (\ref{ext}), for the
range of interest $0.5<c<2, \ 1<x<5$. The result is also insensitive to
the approximation used for the running coupling constant and in fact 
better approximations for the coupling seem to move the values of $S_N$
further away from the lattice data. Finally it should be noted that the 
values $S_N$ also appear to converge as $N$ increases.

The only way to force the values of $S_N$ down closer to the lattice 
data is to choose very low values for the renormalization scale, 
$c \sim 0.05$. Of course this is not only unnatural but increases the
effective value of the coupling contant beyond what one 
would believe is physically reasonable for a perturbative treatment. 
That is by making
an artificially low choice for the renormalization scale, one cannot 
escape the conclusion stated in the abstract of 
large non-perturbative corrections !

\section{Higher Order Corrections}

Due to technical complications the sixth order contribution, 
$\lambda^6$, to the free-energy density has not been calculated 
although an algorithm for it exists \cite{bn}. There is a 
misconception that because that contribution is due to an infinite 
number of topologically distinct diagrams, its value must be very large.
A counter-example is provided by the magnetic screening mass \cite{gpy}, 
which suffers from the same disease but whose approximate 
calculations in the literature show it to be of ordinary 
magnitude \cite{nair}.

Having said that, let us see what is the worst that can happen.
It has been suggested \cite{ellis} that Pade' approximants can be used to estimate the
next term of a truncated perturbation series. That is, after approximating
the truncated series by the ratio of two polynomials, the Pade' approximant
is re-expanded as a power series to estimate the next term in the series. 
Well, why not also use Borel-Pade' approximants for the same purpose ?
Using the fifth order result (\ref{ym}) together with the two-loop
running coupling (\ref{twoloop}), and choosing the central values 
$c=1, \ x=3$, all fifth order Pade and Borel-Pade approximants 
were constructed and then re-expanded to give an estimate of the 
coefficient $f_6$. The largest value obtained was $30,000$ and the
smallest $-30,000$. Note that the fifth order coefficient at $c=1$
is $-800$, so the estimated magnitude of $f_6$ is about $37$ times larger.
Since the coupling $\lambda$ is about $0.2$ at $x=3, c=1$, the total
value of the sixth order contribution to (\ref{ym}) 
is therefore estimated to be almost $8$ times in magnitude compared
to the fifth order contribution. These are big numbers and should 
be expected to cause some damage.  

Using (\ref{rser}) with $f_6=30,000$, the two-loop
coupling (\ref{twoloop}), and solving (\ref{ext}) at the reference 
point $c_0=1, \ x_0=3$ gives $p(6)=19.75$ and $S_6(c=1,x=3)=0.9490$.
Repeating for $f_6 =-30,000$ gives $p(6)=19.5$ and $S_6(c=1,x=3)=0.9489$.
Notice the negligible change in the value of $S_6$ even 
when wildly differing values have been used for $f_6$. Those 
values  should be compared with the fifth order approximant of Fig.(3a),
which gives $S_5(c=1,x=3)=0.9473$. The large estimated sixth order
corrections to the divergent perturbation expansion (\ref{ser}) cause a change of
only $0.002$ to the values of the resummed series, and more importantly 
the shift is upwards, $S_6 > S_5$, preserving the lower order
monotonicity, regardless of the sign of $f_6$.

Kajantie, { \it et.al.} \cite{kaj} 
have suggested that the sixth order contribution, 
$f_6 \lambda^6 $ be of order $10$. For a coupling $\lambda \sim 0.2$, 
this translates into the astronomical value $\pm 156250$ for $f_6$. Solving 
(\ref{ext}) at $c_0=1, x_0$=3 gives $p(6)=19.8$ and $S_6(c=1,x=3)=0.9491$ 
for the positive $f_6$, and $p(6)=19.4$ and $S_6(c=1,x=3)=0.9489$ 
for the negative $f_6$. Despite the anomalously large value of the
sixth order contribution proposed in \cite{kaj}, the conclusion here 
is still $S_6 > S_5$, and an increment of only $0.002$.

To highlight the above result in a more dramatic way, suppose 
the sixth order
coefficient vanishes, $f_6=0$. Then because of the non-trivial 
way the resummation is 
done in (\ref{rser}),
the solution to (\ref{ext}) for $N=6$ will still 
be different from that of $N=5$.
At $c_0=1,x_0=3$ I find  $p(6)=19.5$ and then $S_6(c=1,x=3)=0.9489$ at four 
decimal places, which is almost identical to the values obtained above for
various large values of $f_6$. This sounds incredible but is 
actually not once one
remembers that large corrections to the divergent series (\ref{ser}) do
not translate into large corrections to the resummed series (\ref{rser}).
In fact those large values are suppressed in various ways. Firstly, in the
re-organization of the series in (\ref{rser}), 
less weight is given to higher order corrections. Secondly, the variational
procedure chooses values of $p(N)$ which in this example increase with
$N$, and so suppress further the value of $S_N$.

More understanding of the above results can be obtained through a large
$N$ analysis carried out for the general Eqs.(\ref{rser},\ref{ext})
in \cite{rp2}. It was shown in \cite{rp2} that if $p(N)$ increases 
for the first few values of $N$, then that trend will continue.
Let $c(N) \equiv p(N+1)/p(N)$. In the large $N$ and large $p(N)$ 
limit one can show that \cite{rp2} 
\begin{equation}
{1 \over c(N+1)} = 1- {1 \over c(N)} + {1 \over c(N)^2} \, .
\label{cn}
\end{equation}
One consequence of this is that $c(N+1) < c(N)$ and $c(N) \to 1^{+}$ 
as $N \to \infty$. This is indeed observed for the present problem 
already at low $N$. 
Numerically, (\ref{cn}) too is not a bad approximation 
at small $N$. In fact using 
the values found for $p(N)$ in the last section, one has 
$p(4)/p(3)=7.6/3.2=2.375$. With this as input for $c(3)$, (\ref{cn}) gives 
the estimate $c(4) \sim  1.32234$, to be compared with 
the actual value $p(5)/p(4)=13.1/7.6=1.7$. 
Next using
$c(4)=1.7$ as the exact input, (\ref{cn}) gives $c(5) \sim 1.32201$ and thus
an estimate of $p(6) \sim 1.3 \times 13.1=17.03$. On the other hand, using 
$p(6) \sim 19.5$, as determined by 
various estimates above, gives $c(5)=19.5/13.1=1.49$ and then 
through (\ref{cn}) the estimate $c(6) \sim 1.28$ and 
hence $p(7) \sim 1.28 \times 19.5 =25$.

The Eq.(\ref{cn}) was derived in \cite{rp2} for the case $f_1 \neq 0$.
For the present case where $f_1=0$ one will actually obtain the slightly
more accurate equation
\begin{equation}
{1 \over c(N+1)^2} = 1- {1 \over c(N)} + {1 \over c(N)^3} \, ,
\label{cnn}
\end{equation}
but in the large $N$ limit where $c \to 1^+$ this is clearly
equivalent to (\ref{cn}).

Note that the recursion relations (\ref{cn}, \ref{cnn})
make no explicit reference to the values of the $f_n$ which in the derivation
in \cite{rp2} were assumed to be generic, that is, diverging at most factorially
with $n$. Indeed that fact that various different assumptions
about the value of $f_6$ earlier in this section 
led to essentially the same value for $p(6) \sim 19.5$
supports the $f_n$ independence of 
(\ref{cn}) already at $N \sim 6$. 

From the general analysis in \cite{rp2}, one also
deduces that for large $N$ and large $p(N)$,
the monotonicity $\Delta S_N \equiv S_{N+1}-S_{N} >0$  
is guaranteed by the fact $f_2<0$, and that $\Delta S_N \sim 1/N^3$ 
as $N \to \infty$. 
Since the explicit $N \leq 5$ calculations and the estimated $N=6$ result
already support $c(N)>1$ and large values of $p(N)$ at low $N$,
this suggests  that the continued monotonicity and 
rapid convergence of the $S_N$ is assured by the large $N$ analysis.

\section{Lower Bound and Other Estimates}

From the explicit low $N$ calculations, and the large $N$ analysis,
one concludes that for $N >2$,
\begin{equation}
S_{N} < S_{N+1}
\label{ineq}
\end{equation}
for all $N$, and furthermore, the difference $S_{N+1} -S_{N}$ decreases
as $N$ increases, showing a rapid convergnce of the approximants. 
However, in general, it is not quite correct to say 
that the approximants converge to the total sum of the series
\cite{rp2}. For each $N$, let $p^{\star}(N)$ be the value of $p$ that is optimal,
that is, it is the value which when used in (\ref{rser}) gives the best
estimate of $S$, the total sum of the series.
Define, $S^{\star}_{N} = S_{N}(\lambda,p^{\star}(N))$. 
Then for those $p(N)$ which are positions of global minima one
has by definition,
\begin{equation}
S_{N} \leq S_{N}^{\star}
\label{star}
\end{equation}
It is $S_{N}^{\star}$ which presumably converges to $S$ as $N 
\to \infty$. (This implicitly assumes that the sub-sequence of global minima is infinite: That is, 
given any positive integer $N_0$, there is some $n >N_0$ for which $f_n$
 is positive.) 

Hence if one accepts the two assumptions above, then 
combining (\ref{ineq}) with (\ref{star}),
\begin{equation}
S_{N} \leq S
\end{equation}
for all $N>2$, and one may conclude that the $S_N$ are lower
bounds to the sum of the full perturbation series. 

In particular that conclusion implies that the 
$N=5$ curve in Fig.(5a) is a lower bound on the {\it total perturbative}
contribution to the free-energy density of hot $SU(3)$ theory. 
The statement has three qualifications. Firstly, it involves the 
technical assumptions mentioned above. Secondly, as discussed
before, better approximations to the running coupling can move the bounds,
but it was seen that a $20 \%$ improvement in the coupling shifted the bound 
{\it upwards} by $1 \%$. Thirdly, the bounds shift by $\pm 1 \%$ when the
renormalization scale is varied by a factor of two from its central value 
$\bar{\mu} = 2 \pi T$. Thus it might be more appropriate to call the
bounds as "plausible soft lower bounds" with an uncertainty 
depicted in Fig.(5b). 

Given that the lower bound obtained above involves some unproved technical 
assumptions, it is useful to compare the above results with those obtained using
different resummation schemes for the divergent series (\ref{ser}).
I briefly state here the main results obtained using a 
Borel-Pade' resummation of (\ref{ser}), with the two-loop approximation
for the coupling (\ref{twoloop}) and the central value $c=1$. 
The approximants will be denoted
as $[P,Q]$, referring of course to the particular Pade' 
approximant used for
the partial Borel series (\ref{borel}) constructed from (\ref{ser}).
The only approximants which did not develop poles and which gave a 
resummed value below one in the temperature range $2<x<5$
were $[1,2], \ [2,1]$ and $[2,2]$.
These are displayed in Fig.(6). The $[3,2]$ and $[4,1]$ approximants
did not develop poles but gave a value above one. If the approximants 
which developed a pole are defined through a principal value prescription,
then the lowest value was given by $[2,3]$: $0.91 \to 0.94$ as $x$ increased from
$2 \to 5$. The $[1,3]$ and $[1,4]$ approximants gave values above $0.98$
in the range of interest while $[3,1]$  gave a value above one.

Thus in the Borel-Pade' method, the minimum estimate for 
the fifth order resummed series is given by 
the principal value regulated $[2,3]$.
The highest values were all above one. If one keeps only the fifth order 
estimates below one (thus giving a very conservative lower value), then 
the average of the $[2,3]$ and $[1,4]$ is greater than $0.94$ for
the entire range $2<x<5$. At $x=3$ the estimates are $0.95 \pm 0.04$. 
Of course including also the values above one would push this average
higher. Clearly the Borel-Pade' estimates are comparable to the 
bounds obtained using the resummation technique 
of Sec.(2) and should reassure some readers about the novel resummation
used here.

For completeness, I mention an alternative way of thinking about
divergent series such as (\ref{ser}). For QCD at zero temperature,
a paradox is that one-loop results give remarkable agreement with
experimental data even when the energy scale is relatively low.
As
the running coupling is then large it is not obvious why higher-loop
perturbative corrections are suppressed. It has been 
suggested \cite{sterman,west} that the explanation
might lie in the probabale asymptotic nature of the QCD 
perturbation series. Recall that in an asymptotic series 
the best estimate of the full sum, {\it at a
given value of the coupling}, is obtained when only an  
optimal number of terms is kept and the rest discarded 
(even if they are large).
Thus if one knew the general behaviour, at least at large order, of
the series (\ref{ser}) and assumed that it was asymptotic, then one
could have obtained a reasonable estimate of the full sum 
by simply adding the optimal first few terms. 
What has been done in the previous 
sections, and this is what various resummation schemes try to do,
is to instead sum up the whole series to get an even better estimate 
of the total perturbation series (and this has the greater
advantage of giving a good result for a large range of couplings). 
Also note that thinking of
(\ref{ser}) as an asymptotic series does not say anything about explicit  
non-perturbative corrections \cite{sterman,west}.

\section{Non-Perturbative Corrections}

The total perturbative contribution to the 
free-energy density of $SU(3)$ gauge theory has been argued to be 
close to, or above, the $N=5$ curve in Fig.(5a). A residual uncertainty 
that could lower the curve of Fig.(5a) is the
exact value of the renormalization scale. For a natural range of 
parameters, the lower curve in Fig.(5b) is the result.
On the other hand 
the full result as given by lattice simulations is shown in Fig.(1).
Lattice errors have been stated to be under $5\%$ \cite{latt}. Taken together,
the conclusion appears inescapable : {\it Even at temperatures  
a few times above 
the transition temperature, there are large negative non-perturbative
contributions to the free-energy density}. For example, at $T= 3 T_c \sim 700 MeV $,
the lattice results for the normalised free-energy density are $0.8 \pm 0.04$
while the lower bound on the perturbative contribution is $0.947 \pm 0.007$,
implying a minimum non-perturbative correction of $10 \%$ (and as high as $20 \%$).   

Thus an answer has been given to the questions raised in the introduction.
The deviation of the lattice data from the ideal gas value is apparently
caused mainly by non-perturbative corrections, 
with perturbative corrections accounting 
for a much smaller amount. At $T \sim 3 T_c$ the relative contributions
are $ \sim 15\% $ and $\sim 5\% $.
 
I speculate now on possible sources of the non-perturbative corrections.
Firstly there are the familiar instantons, already present in the classical 
action, and which contribute terms of the
order $e^{-1/\lambda^2}$. Secondly there are the magnetic 
monopoles. There is by now
overwhelming evidence that confinement at zero temperature 
is caused by the t'Hooft-Mandelstam mechanism 
of condensing monopoles (the dual superconducting vacuum).
Thus it is possible that the monopole condensate has not 
completely melted above the critical temperature. 
Note that since the classical theory does not support finite energy monopoles,
these must be of quantum origin, and so their contribution might be 
larger than those of the instantons.

In fact contributions which are exponentially small but much larger 
than those of the instantons are suggested by the Borel resummation itself.  
It is known that Yang-Mills theories are not Borel summable \cite{qcd,ben}.
That is, the function $B(z)$ has singularities for positive $z$, making the 
Borel integral ill-defined. One can nevertheless define the sum of the
perturbation series using the Borel integral if a prescription is
used to handle the singularities. It is generally believed that the prescription
dependent ambiguity disappears when explicit non-perturbative contributions
are taken into account for the physical quantity in question. 
Indeed the nature of singularity itself suggests the form for the
non-perturbative contribution. If there is a pole at $z=q$, then the
non-perturbative contribution will be of the form $\sim e^{-q/\lambda}$,
which is larger than the instanton contribution for small $\lambda$.
An explicit mathematical model which illustrates the interplay between
Borel non-summability and non-perturbative contributions 
has been given in \cite{rp2}.

Notice that the non-perturbative corrections suggested by the Borel 
method at non-zero temperature are very different from those at zero
temperature. In the latter case the expansion parameter is $g^2$ and so the
contribution is $\sim e^{-q/g^2}$, which translates into a power suppressed
contribution $\sim 1/(Q)^b$ 
when $g^2$ is replaced by the running coupling 
$\sim 1/\ln(Q/\Lambda)$. 
In cases where the physical quantity can also be analysed using the 
operator-product expansion (OPE), these power suppressed contributions to
perturbative results correspond in the OPE picture to vacuum condensates
\cite{qcd,ben}.

At nonzero temperature, since the natural expansion parameter is 
$\lambda = \sqrt{g^2 /4 \pi^2} \sim 1/\sqrt{\ln(T/\Lambda)}$, one does not 
get a simple power suppression from $e^{-1/\lambda}$. Nevertheless, the
analogy with zero-temperature results suggests that such 
contributions might be
due to some condensates. Thus the conventional condensates discussed for 
example in \cite{kaj}
are plausibly part of the non-perturbative contributions. 
The form however suggests even more novel condensates. 
These might be, for example, those of  
DeTar \cite{detar} or Pisarski \cite{pis}.

It is worth noting an explicit instance of a theory displaying 
exponentially 
small non-perturbative effects which are larger than those 
due to standard solitons. In fundamental 
string theory where the coupling is $g$, there are the usual 
solitons but there are also novel "D-instantons"
which give a larger contribution $e^{-1/g}$ \cite{dbrane}. 
I also mention in passing the
recurrent and intriguing relationship between gauge theories and strings
\cite{holo}
which leads one to wonder whether that is a possible route 
to understanding the non-perturbative structure of hot gauge theories.

Using the $N=5$ curve of Fig.(5a) as a reasonable estimate of the
full perturbative result, 
and assuming a nonperturbative component of the form
\begin{equation}
S_{np} = {A \over \lambda} \ e^{- q/\lambda} \, ,
\label{bag}
\end{equation}
as suggested by the Borel method, one can determine 
the constants $A$ and $q$ by comparing the lattice data of Fig.(1)
with the perturbative result. In \cite{rp1} it was shown that,
\begin{equation}
S_{latt} = S_{pert} - {1 \over \lambda(x)} \ e^{8.7 - 2.62/\lambda(x)} \ ,
\label{bbag}
\end{equation}
where $S_{latt}$ represents the lattice data for the free-energy, 
and $S_{pert}$ the resummed perturbative result, 
both normalized with respect to the ideal gas value, and $\lambda(x)$
is given by (\ref{twoloop}) at $c=1$.
Eq.(\ref{bbag}) is a phenomenological equation
of state for the free-energy which generalises the usual
discussions in the literature where the
second term on the right-hand-side
of (\ref{bbag}) is called a 'bag constant'. In this case 
the `constant' is really
temperature dependent and represents a non-perturbative contribution to the
free-energy that
vanishes at infinite temperature.
It is important to note that the non-perturbative contribution is negative,
since the perturbative result is above the full lattice data,
and thus consistent with the usual interpretation in the literature. 

So far the discussion has implicitly assumed an additive picture of
perturbative and non-perturbative contributions, with both
components clearly distinguished. It might be that in reality 
the best 
description of the high-temperature phase is in terms of 
completely novel structures \cite{detar,pis}. In that case 
a forced expansion of those alternatives 
about $\lambda=0$ must give something like
(\ref{bbag}) and the subsequent distinction between perturbative and 
non-perturbative contributions. The mathematical
toy model of \cite{rp2} illustrates this.

Within the framework of this paper, one can distinguish three versions
of the popular concept of  "quasi-particle". Firstly there are the 
"perturbative quasi-particles" which are deformations of the
gluon formed by a particular reorganisation of the 
perturbative Feynman diagram expansion. 
The results above suggest that if all the contributions of 
such quasi-particles 
to the free-energy are added up, the net result will lie above 
the lattice data, and only
truly non-perturbative contributions, as defined in Sec.(1), 
may give the final agreement. Secondly there are the "nonperturbative
quasi-particles", which are excitations about the nontrivial thermal 
vacuum that includes condensates, and so forth. 
Currently there is not sufficient control over the theory to construct these
objects. Finally there are the "phenomenological quasiparticles" 
which simply aim
to give numerical agreement with the lattice data within a 
simple {\it ansatz}.
The ultimate justification for these phenomological constructs 
must surely come from the
"nonperturbative quasiparticles".

\section{QCD}

In this section lower bounds (within assumptions similar to those made
previously) are obtained for 
the perturbative free-energy density of hot $SU(3)$ coupled to $N_f$ 
flavours of fundamental fermions. As the essential features 
are very similar 
to the pure gauge case, the discussion here will be brief. The 
fifth order perturbative 
results in the $\overline{MS}$ scheme 
can be read off from the landmark papers \cite{free}. The approximate
two-loop coupling that is used here is given by 
\begin{equation}
\lambda(c,x) = {1 \over \sqrt{4 \pi^2 \beta_0 L(c,x)}} \left( 1-{\beta_1 \over 2 \beta_{0}^{2}}
{\ln(L(c,x)) \over L(c,x)} \right) \,
\label{newloop}
\end{equation}
with 
\begin{eqnarray}
\beta_0 &=& {11 N_c - 2 N_f \over 48 \pi^2} \, ,\\
\beta_1 &=& {1 \over 3(4 \pi)^4 } (34 N_{c}^{2} -13N_c N_f + 3N_f/N_c) \, .
\end{eqnarray}
Following \cite{blaz} I also assume a relative  $N_f$ independence of 
$L(c,x)$, and thus use for it the same expression as used in (\ref{oneloop}).
The extremization condition (\ref{ext})
is solved at the reference point $c_0=1, \ x_0=3$ for $N_c=3$ and 
$1 \leq N_f \leq 6$. 
The results obtained are all extremely 
similar: In each case the convergence of the $S_N$ is monotonic and rapid
as in the pure glue case in Fig(5a). For this reason only the $N=5$ curves are
displayed in Fig.(7) for the various number of flavours. For comparison 
the pure glue result ($N_f=0$) is also included. (Each curve
has been normalized with respect to the ideal gas value for that number  
of flavours.)

The $N=5$ curves in Fig.(7) can be taken as plausible lower bounds, or estimates,
to the total perturbative free
energy density of QCD with $N_f$ fundamental fermions. 
Lattice results for QCD contain large systematic 
errors compared to those for $SU(3)$ and so a precise comaprison is 
not possible.
After making some assumptions about the size of the systematic errors,
the authors in Ref.\cite{latt2} determine that for $N_f =2$ the
free-energy density lies about $15-20 \% $ below the ideal gas limit. 
This is similar to the case of pure $SU(3)$. Comparing this lattice
estimate with the estimate on the perturbative result in 
Fig.(7) one is again led to suggest that
there are large non-perturbative corrections to the naive picture
of a weakly interacting quark-gluon plasma. 

Of course, given the physical relevance of QCD, it would be 
preferable to have more precise numbers from the
lattice, and especially for other values of $N_f$. However it seems that
a non-perturbative component of $10-15\% $ at temperatures a few times $T_c$ 
is likely to be generic.

When the draft of this paper was complete, I came across \cite{gupta}
which gives for $N_f=2$ a $T_c/\Lambda_{\overline{MS}} \sim 0.5$,
a factor of two lower than that for the pure gauge theory. This has the 
consequence that $c$ in $L(c,x)$ should be replaced by about $c/2$.
However as the reader can surmise by now, this has hardly any impact on the 
results above, for this is equivalent to shifting the renormalization 
scale $c$ by a factor of two, which as we have seen causes only 
a $1\% $ shift of the curves. In any case this serves to remind 
that the lattice 
results for $N_f$ are in a state of flux.

\section{SU(N)$_c$}

Define 
\begin{equation}
\lambda(N_c) = \left({N_c \over 3}\right)^{1/2} \  \left({\alpha \over \pi}\right)^{1/2} \, .
\label{universe}
\end{equation}
Then the free-energy density of pure $SU(N_c)$ theory up to fifth
order \cite{free,bn} is given by the expression (\ref{ym}), 
with $(\alpha / \pi)^{1/2}$
replaced everywhere (including inside the logs) 
by $\lambda(N_c)$. Thus there is no explicit $N_c$ dependence
of the free-energy density when written in terms of $\lambda(N_c)$. 

To 
examine the $N_c$ dependence of the new coupling (\ref{universe}), 
consider the
approximate two-loop running coupling
given by  
\begin{equation}
\left({\alpha(T) \over \pi}\right)^{1/2} = {1 \over \sqrt{4 \pi^2 \beta_0 L(T,\overline{\Lambda})}} \left( 1-{\beta_1 \over 2 \beta_{0}^{2}}
{\ln(L(T,\overline{\Lambda})) \over L(T,\overline{\Lambda})} \right) \,
\label{newbeta}
\end{equation}
with
\begin{eqnarray}
L(T,\overline{\Lambda}) & \equiv& 2 \ln \left({2 c' \pi T \over \overline{\Lambda} }\right)  \, , \\
\beta_0 &=& {11 N_c  \over 48 \pi^2} \, , \\
\beta_1 &=& {1 \over 3(4 \pi)^4 } (34 N_{c}^{2}) \, ,
\end{eqnarray}
and with $\overline{\Lambda}= \overline{\Lambda}(N_c)$ 
the $SU(N_c)$ gauge theory scale parameter in the $\overline{MS}$ scheme. 
In $L$, the constant $c'$ 
is $\bar{\mu} / 2 \pi T$. Comparing the various equations, one 
comes to the remarkable conclusion 
that the new running coupling $\lambda(N_c, T)$ will be independent 
of $N_c$ {\it if} the $\overline{MS}$ scale parameter $\Lambda(N_c)$
is itself independent of $N_c$ when expressed in terms of some physical length scale. 
By comparing some data for $N_c =2,3,4$,
Teper \cite{teper} has concluded that this is indeed the case. 

Therefore, accepting the result of \cite{teper}, one deduces that 
the $N=5$ curve in Fig.(5a) is a plausible lower bound, or estimate,
to the total perturbative free
energy density of hot $SU(N_c)$ theory when the x-axis is interpreted as 
$T / \overline{\Lambda}$ instead of $T/ T_c$. This is then a universal
relation
(at least for low $N_c$), and one suspects that the corresponding full
lattice results might also obey a universal curve, thus leading to the guess
that the non-perturbative component of an $SU(N_c)$ plasma
is about $10-15 \%$ for temperatures a few times 
above the critical temperature.

\section{QED}
Though the fine structure 
constant $\alpha$ of QED is small at everyday energies, 
it is interesting to 
consider super-high temperatures where it will be large.
The free-energy density of massless QED at temperature ($T$), has been
computed up to order $\alpha^{5/2}$ in Refs.\cite{qed,free,bn,an}. 
Denoting as usual $\lambda= (\alpha / \pi)^{1/2}$,
the normalised free-energy density at the $\overline{MS}$
renormalisation scale $\bar{\mu}=2 \pi T$, is given by \cite{qed,free}
\begin{equation}
{F/F_0} = 1 - 1.13636 \lambda^2 +  2.09946 \lambda^3 +  0.488875 \lambda^4 -  6.34112 \lambda^5 \, .
\label{free}
\end{equation}
where $F_0= 11 \pi^2 T^4 / 180$ is the free-energy 
density of a non-interacting plasma of electrons, positrons and photons.
Figure (8a) shows the plot of (\ref{free}) at different orders.
The series diverges at large coupling (super-high temperatures),
exhibiting a behaviour similar to that of Yang-Mills theory at low-temperatures.
The convergence at large coupling
can be improved by using
the resummation technique (\ref{rser}-\ref{ext}). Using the coefficients
from (\ref{free}), the solutions of (\ref{ext}) at the reference value
$\lambda_0 =0.5$ are (minima): $ p(3)=0.7, \ p(4)=1.75, \ p(5)=3$.

The resummed series, with its much improved convergence, 
is shown in Fig.(8b). The $N=5$ curve can be taken as a lower bound
to the full perturbative result. 
If one assumes that the potential 
non-perturbative contributions lower the perturbative
result, as happens in QCD, or are very small in magnitude,
then one may conclude from Fig.(8b) that
super-hot QED undergoes a phase transition.
This speculated high-temperaure phase of QED
might then be analogous to the low-temperature phase
of QCD with various bound states.
Or, it might resemble the alternative picture of low-energy 
QCD: that of flux-tubes \cite{gold}. It is unfortunate that no
lattice or other non-perturbative information is currently 
available about the
high-temperature phase of QED.

\section{Conclusion}

The phrase 'non-perturbative' is used often and loosely 
with regard to field theories at 
non-zero temperature. This has caused a great deal of semantic confusion 
and misunderstanding. For the purpose of uncovering
the cause of the deviation
of the result in Fig.(1) from the ideal gas value, it has been proposed
to term 'perturbative' all power like (modulo logarithms) contributions 
to the free-energy density. Such perturbative contributions follow
from the usual Feynman diagram expansion of the parition function
around zero coupling. 

For $SU(3)$ gauge theory a plausible 
lower bound was obtained on the totality of
such perturbative contributions to the free-energy density. 
The derivation of that lower bound using the variational conformal map
involved some technical assumptions, 
and so one may instead wish to consider it only 
as an estimate of the total perturbative
contribution. The estimate is comparable to that obtained using Pade'
or Borel-Pade' resummation methods and lies significantly
higher than the full lattice result, thus suggesting that
large and truly non-perturbative corrections exist. As discussed 
in Sec.(6), these
non-perturbative corrections might include the usual instantons, magnetic
monopoles, the usual condensates, and perhaps also more novel condensates
and extended structures as suggested by the Borel form $e^{-1/\lambda}$ 
of the non-perturbative contributions. As to which of these possibilities
dominates is an interesting question left for future work. 

The equation of state for hot 
$SU(3)$ can be summarized by the phenomenological relation \cite{rp1}
\begin{equation}
S_{latt} = S_{pert} - {1 \over \lambda(x)} \ e^{8.7 - 2.62/\lambda(x)} \ ,
\label{bbag2}
\end{equation}
where $S_{latt}$ represents the lattice data for the free-energy, 
and $S_{pert}$ the resummed perturbative result, 
both normalized with respect to the ideal gas value, and where $\lambda$ is
the temperature dependent coupling (\ref{twoloop}) at $c=1$. 
There is a slight ambiguity in the estimate of the magnitude of 
non-perturbative corrections
coming from the residual renormalization scale ambiguity of the resummed
perturbative results. For the natural range $\pi T < \bar{\mu} < 4 \pi T$,
the ambiguity is less than one percent. Such an ambiguity between the
perturbative and non-perturbative components is understandable, as only the
full physical quantity can be demanded to be scale independent, 
and not separately its perturbative and non-perturbative components, though 
the latter simplifying assumption is often made.

By choosing an anomalously
low value for the scale $\bar{\mu}$ 
one can fit the purely perturbative results to the lattice
data but at the price of a very large effective coupling constant, 
and thus an unsuppresed contribution from common place objects like
instantons. Therefore this route does not offer an escape from the 
conclusion of large non-perturbative corrections.

The results of this paper do not imply that the description 
of the free-energy density must be as in (\ref{bbag2}). Rather, 
completely
novel descriptions in terms of various extended structures are possible
\cite{detar, pis}, but in a forced expansion of those alternative 
descriptions about $g=0$, one must recover
something like (\ref{bbag}). On the other hand, the results 
here {\it do} imply
that an accounting of the lattice free-energy density in terms of
a subset of Feynman diagrams of the perturbative thermal vacuum 
might miss some essential physics.

Being restricted to an analysis of the free-energy density, the results
here of course do not imply that every observable must have 
large non-perturbative contributions, but it is likely that this is true
for most of the bulk thermodynamic quantities.

An analysis was also carried out for generalized QCD with $N_f$ quarks. 
Currently the lattice data is only approximate and so the conclusions 
are less definitive. However accepting the estimates in \cite{latt2}, the
conclusion is the same as before: There appear to be large non-perturbative 
corrections
to the free-energy density of hot $SU(3)$ gauge theory 
coupled to $N_f$ quarks. 
Thus in particular, "quark-gluon plasma" seems to be 
an incomplete description
even at temperatures several times above the transition temperature.

A simple relation was noted for the perturbative 
free-energy density of $SU(N_c)$ gauge theory in Sec.(8). That result,
combined with the results for generalised QCD in Sec.(7), 
and the available lattice data,
leads one to a universality {\it conjecture}: For the 
high-temperature ($T \geq 2 T_c$)  phase of
$SU(N_c)$ gauge theory coupled to $N_f$
fundamental quarks, and for all moderate values of $N_c$ and $N_f$, 
at most $5 \%$ of the deviation of 
the free-energy density from the ideal gas 
value is due to perturbative effects 
while non-perturbative effects  contribute a larger $10-15 \%$.

It has been speculated in Sec.(9) that QED might 
have an interesting high-temperature phase.

Finally, the methods of this paper might be of
some use for the study of supersymmetric theories at non-zero temperature,
a topic of interest in recent developments \cite{holo}. \\

\vspace{0.3in}
\noindent
{\large \bf Acknowledgements}: I thank B. Choudhary, C. Coriano$'$,
A. Goldhaber, U. Parwani, I. Parwani, S. Pola, J.S. Prakash, D. Saldhana,
S. Saldhana, and  P. Van Nieuwenhuizen for their hospitality during
the course of this work.\\

\newpage

{\bf Figure Captions}

Figure 1: 
Mean lattice results for the free-energy density of hot $SU(3)$ gauge
theory from Ref.\cite{latt}. Here $S_{latt}$ refers to the free-energy
divided by the free-energy of an ideal gas of gluons.
\\

Figure 2:
The divergent perturbative free-energy density of $SU(3)$ gauge theory
given in Eq.(\ref{ym}). Starting from the lowest curve at 
$(\alpha/2\pi)^{0.5} =0.24$,
one has $N=2,5,3,4$.
\\

Figure (3a):
The resummed perturbative 
free-energy density of hot $SU(3)$ gauge theory
for $N=3,4$ and $5$, using a one-loop running coupling, the
reference values $c_0=1, \ x_0=1$, and the renormalization scale $c=1$.
The curves move upwards as $N$ increases.
\\

Figure (3b):
The fifth order resummed perturbative free-energy density of 
Fig.(3a) at three different renormalization scales, 
$c=0.5, \ 1$ and $2$. The free-energy density 
increases with increasing $c$.
\\

Figure 4:
The one-loop (upper curve) and approximate two-loop running couplings for
$SU(3)$ gauge theory at the renormalization scale $c=1$. 
\\

Figure (5a):
The resummed perturbative 
free-energy density of hot $SU(3)$ gauge theory
for $N=3,4$ and $5$, using a two-loop running coupling, the
reference values $c_0=1, \ x_0=3$, and the renormalization scale $c=1$.
The curves move upwards as $N$ increases from $3$ to $5$.
\\

Figure (5b): 
The fifth order resummed perturbative free-energy density of 
Fig.(5a) at three different renormalization scales, 
$c=0.5, \ 1$ and $2$. The free-energy density 
increases with increasing $c$.
\\

Figure 6:
The $[1,2], \ [2,1]$ and $[2,2]$ Borel-Pade' approximants to the
perturbative free-energy density, with a two-loop running coupling,
and the renormalization scale $c=1$.
Starting with the lowest curve at $x=5$, 
one has $[2,1], \ [1,2], \ [2,2]$.
\\

Figure 7:
The fifth order resummed perturbative 
free-energy denisty of $SU(3)$ gauge theory
coupled to $N_f$ fermions as discussed in Sec.(7). Starting 
from below at $x=5$, the curves label $N_f=6,5,4,3,0,2,1$.

Figure (8a):
The divergent perturbative free-energy density of QED, 
given in Sec.(9). Starting from the lowest
curve and moving upwards, one has $N=2, \ 5, \ 3, \ 4$.  
\\

Figure (8b):
The resummed perturbative free-energy density of QED.
The curves move upwards as $N$ increases from $3$ to $5$. 
\\

\newpage

\input{epsf.sty}

\epsfbox{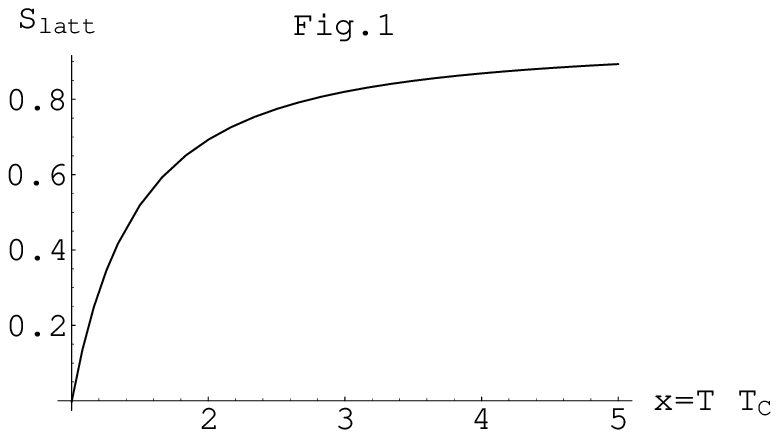}
\vspace{0.5in}

\epsfbox{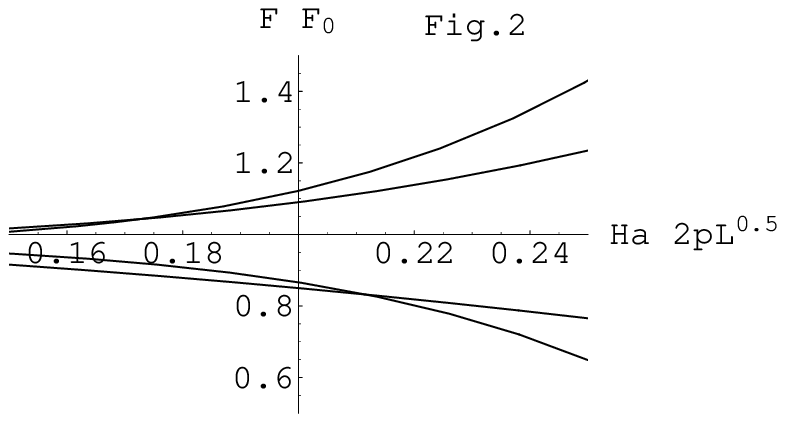}
\vspace{0.5in}

\epsfbox{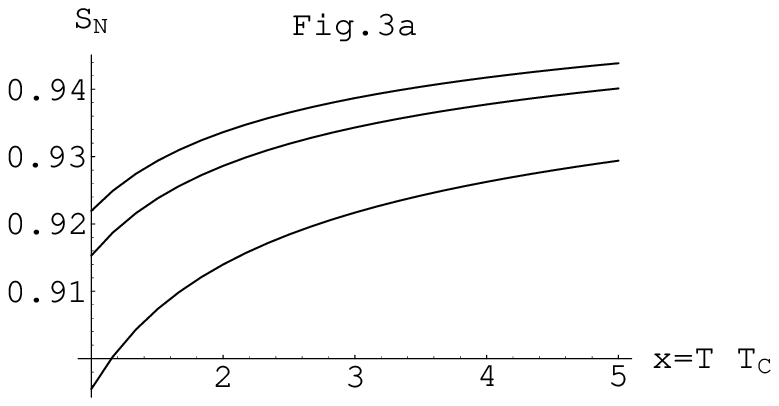}
\vspace{0.5in}

\epsfbox{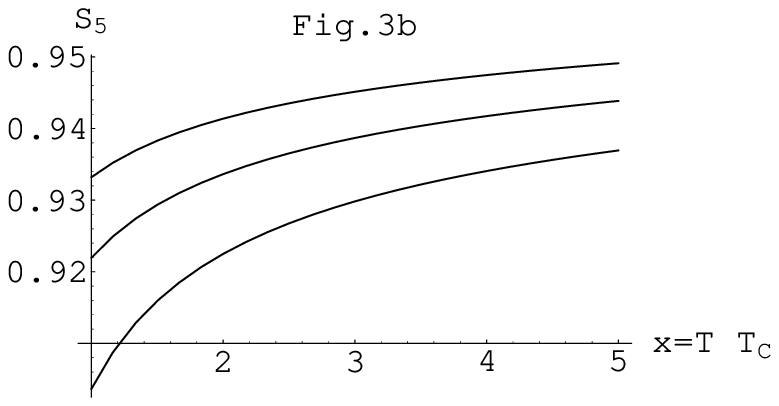}
\vspace{0.5in}

\epsfbox{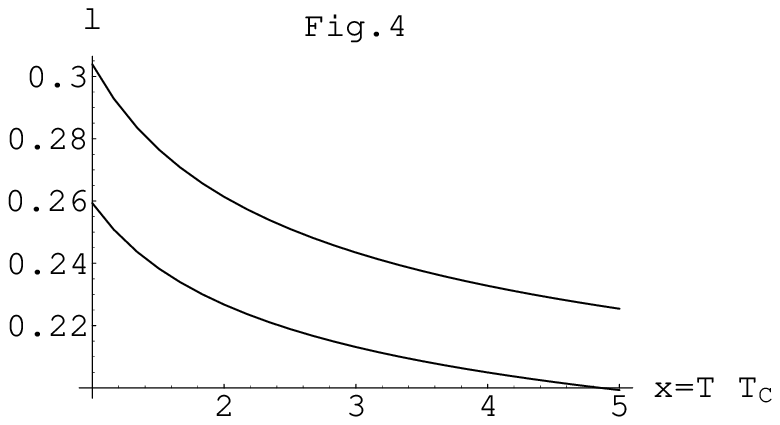}
\vspace{0.5in}

\epsfbox{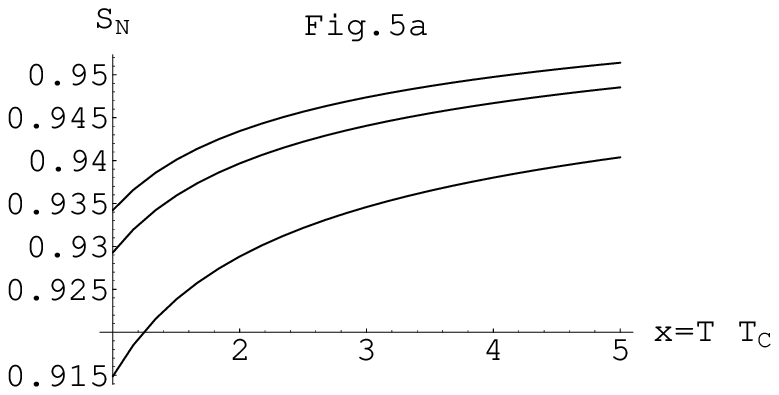}
\vspace{0.5in}

\epsfbox{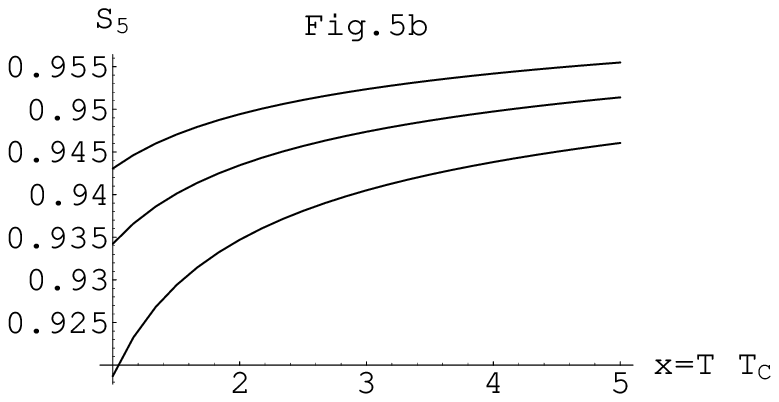}
\vspace{0.5in}

\epsfbox{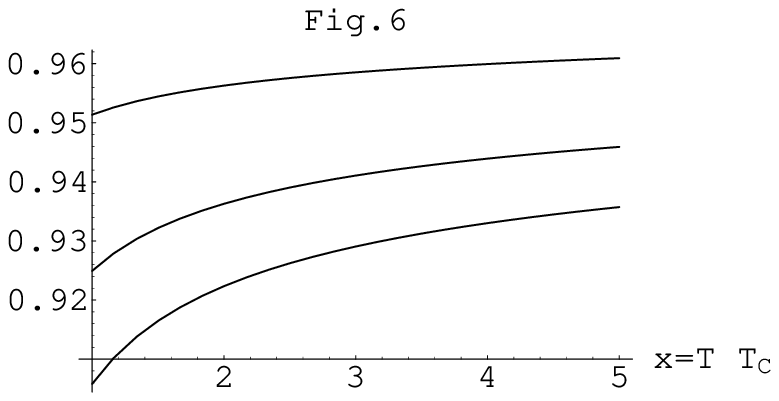}
\vspace{0.5in}

\epsfbox{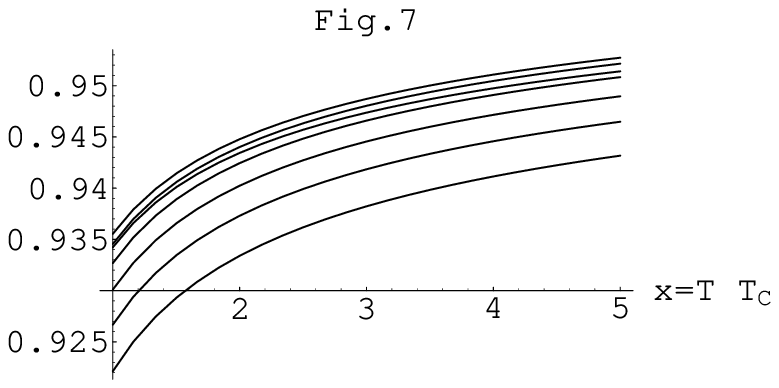}
\vspace{0.5in}

\epsfbox{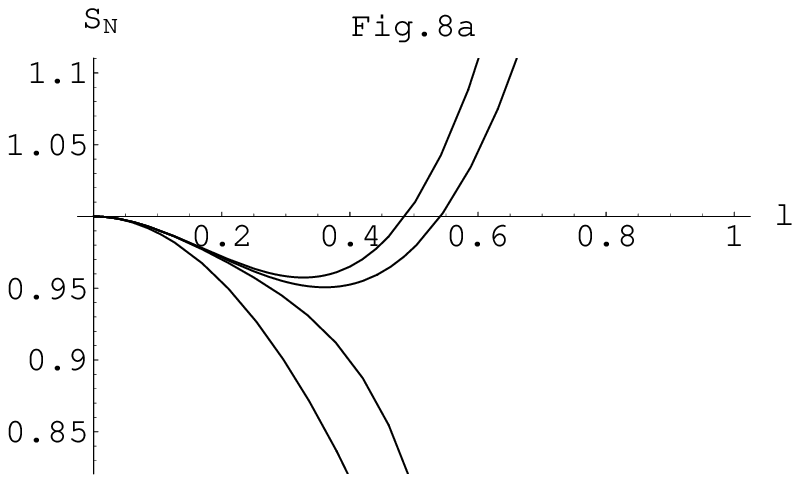}
\vspace{0.5in}

\epsfbox{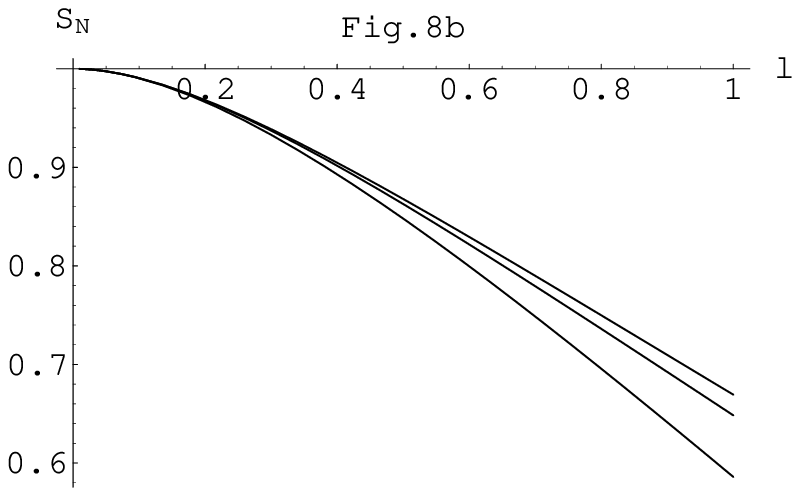}

\end{document}